\documentclass[11pt]{article}
\usepackage[top=1in,bottom=1in,left=1in,right=1in]{geometry}

\usepackage[utf8]{inputenc} 
\usepackage[T1]{fontenc}    
\usepackage{hyperref}       
\usepackage{url}            
\usepackage{booktabs}       
\usepackage{amsmath}
\usepackage{amssymb}
\usepackage{amsfonts}
\usepackage{amstext}
\usepackage{amsthm}
\usepackage{nicefrac}       
\usepackage{microtype}      
\usepackage{lipsum}
\usepackage{graphicx}       
\usepackage{xcolor}
\usepackage{multirow} 
\usepackage[round]{natbib}
\usepackage{svg}
\usepackage{adjustbox}
\usepackage{soul}

\usepackage{comment,multirow,stmaryrd,subfigure,empheq,caption,stackrel,mathabx,mathtools,algorithm2e,amsthm,rotating}

\title{Online Fault Detection and Classification of Chemical Process Systems Leveraging Statistical Process Control and Riemannian Geometric Analysis}

\author{
  Alireza Miraliakbar$^{1,2}$ \and Fangyuan Ma$^{1,3}$ \and Zheyu Jiang$^{1 \star}$\\
}

\date{
    \normalsize $^1$School of Chemical Engineering, Oklahoma State University, Stillwater, OK, 74078 USA\\
    $^2$Department of Chemical and Biomolecular Engineering, University of Connecticut, Storrs, CT, 06269 USA\\
    $^3$College of Chemical Engineering, Beijing University of Chemical Technology, Beijing, 100029 China
}

\begin{document}
\maketitle
\vspace{-1em}
\noindent Corresponding author: \texttt{zjiang@okstate.edu}$^\star$

\begin{abstract}
In this work, we study an integrated fault detection and classification framework called FARM for fast, accurate, and robust online chemical process monitoring. The FARM framework integrates the latest advancements in statistical process control (SPC) for monitoring nonparametric and heterogeneous data streams with novel data analysis approaches based on Riemannian geometry together in a hierarchical framework for online process monitoring. We conduct a systematic evaluation of the FARM monitoring framework using the Tennessee Eastman Process (TEP) dataset. Results show that FARM performs competitively against state-of-the-art process monitoring algorithms by achieving a good balance among fault detection rate (FDR), fault detection speed (FDS), and false alarm rate (FAR). Specifically, FARM achieved an average FDR of 96.97\% while also outperforming benchmark methods in successfully detecting hard-to-detect faults that are previously known, including Faults 3, 9 and 15, with FDRs being 97.08\%, 96.30\% and 95.99\%, respectively. In terms of FAR, our FARM framework allows practitioners to customize their choice of FAR, thereby offering great flexibility. Moreover, we report a significant improvement in average fault classification accuracy during online monitoring from 61\% to 82\% when leveraging Riemannian geometric analysis, and further to 84.5\% when incorporating additional features from SPC. This illustrates the synergistic effect of integrating fault detection and classification in a holistic, hierarchical monitoring framework.

\textit{Keywords}: Process monitoring, Fault detection, Fault classification, Riemannian geometry, Statistical process control

\end{abstract}

\section{Introduction}

Enabled by continuous advancements and integration of sensor technologies and digital infrastructure (e.g., the distributed control system), modern chemical plants have been heavily relying on effective and reliable online process monitoring to enhance product quality, reduce operating costs and environmental footprint, and ensure process safety as they continue to expand in scale and complexity \cite{psereview}. Massive arrays of online process data streams, which monitor the plant's equipment performance, manufacturing processes, and mass, energy, and information flows, are continuously produced by numerous sensors and can be leveraged for process monitoring and control \cite{monitoringreview}. These online data streams are often nonparametric (i.e., data streams do not necessarily follow any specific distribution) and heterogeneous (i.e., data streams do not necessarily follow the same distribution), thereby posing challenges to effective chemical process monitoring, such as early fault detection and classification.

Over the past decades, a number of algorithmic methods have been developed to utilize the large volumes of historical and online sensor data for reliable online process monitoring. Among them, dimensionality reduction approaches, such as principal component analysis (PCA) \cite{pca} and partial least squares regression (PLS) \cite{pls1,pls2}, have received the most attention in the literature \cite{chiang,FEZAI20181}. While traditional PCA and PLS use linear orthogonal transformations to extract data features, modern chemical processes are typically nonlinear \cite{fangyuan}. Thus, nonlinear variations of these methods, such as locally weighted and kernel PCA or PLS \cite{kpca,locallyweighted}, have been proposed. Nevertheless, these dimensionality reduction-based approaches assume that the statistics used to represent the in-control profiles must also span the subspace defining the out-of-control states \cite{Woodall01072004}. In other words, if one monitors only a few features (e.g., principal components) obtained from historical in-control process data, one should ensure there is not a profile shift in some otherwise undetectable direction during online monitoring. However, in chemical process monitoring, process dynamics can become quite complex and out-of-control states (i.e., process anomalies or faults) cannot be fully enumerated or anticipated a priori. Also, the detection result obtained from these dimensionality reduction-based methods are usually difficult to interpret by operators and process engineers because the features are in the reduced space, which does not have a one-to-one mapping to the original big data sources. In addition, since the number of possible fault scenarios can be quite large, monitoring only the most significant subset of features can cause significant error, as the fault may not be noticeable in the selected features. Furthermore, most existing dimensionality reduction-based methods inherently assume that the full observations of all data streams follow a well-known distribution (e.g., normal distribution). Little work has been undertaken on validating the performance of these dimensionality reduction methods on monitoring nonparametric and heterogeneous online data streams.

More recently, advancements in machine learning (ML) have stimulated the development of various chemical process monitoring algorithms based on supervised methods such as support vector machine \cite{svm_1,svm_2}, convolutional neural network, \cite{cnn_1,cnn_2} and recurrent neural network \cite{rnn_1,rnn_2}, as well as unsupervised methods such as autoencoder \cite{ae_1}. Nevertheless, machine learning-based process monitoring techniques still face issues such as overfitting. Also, the process monitoring performance of supervised ML-based methods relies on the availability of a large amount of labeled data, especially faulty data, which are typically limited and hard to acquire in practice. Moreover, ML-based methods do not scale well for new fault scenarios that have not been encountered before.

The limitations of conventional dimensionality reduction and ML-based process monitoring methods motivate us to explore alternative approaches to more effectively conduct fault detection and classification. In terms of alternative methods for fault detection, we have caught attention to latest advancements in statistical process control (SPC) techniques \cite{montgomery2008}. Univariate control charts such as Shewhart chart \cite{Shewhart}, cumulative sum (CUSUM) \cite{cusum}, and EWMA \cite{Roberts01081959} are proven, effective tools that have a solid theoretical foundation and have been extensively studied and used for process monitoring for decades \cite{shewhart_1, ewma_1, ewma_2, cusum_1, cusum_3}. Furthermore, one distinct feature of these SPC techniques is that users can specify their desired false alarm rate (Type I error), thereby offering plant operators the much-needed guarantee and flexibility to safely adjust the alarm sensitivity based on their understanding of the process. Without loss of generality, we choose the CUSUM-based approach as our baseline as these univariate control charts have similar features. When it comes to modeling several data streams simultaneously, one needs to extend the univariate CUSUM procedure to multivariate ones by first constructing the local statistic for each individual data stream, followed by combining these local statistics together somehow into a single global monitoring statistic. Assuming the local statistic of each data stream follows a normal distribution, \citet{mei} proposed a global monitoring scheme based on the sum of the largest $r$ local statistics. \citet{TARTAKOVSKY2006252} proposed using the maximum of local statistics to construct the global monitoring statistic. However, these earlier multivariate CUSUM models are parametric and homogeneous and thus are quite restrictive. And we suspect that these drawbacks have hindered SPC techniques from attracting more attention among industrial practitioners and the process systems engineering (PSE) community.

To address these drawbacks, nonparametric multivariate CUSUM procedures have been recently developed \cite{Qiu30042008,Bakir01052006}. Most of these nonparametric procedures are based on the idea of monitoring the data stream indices in the ascending-order and descending-order ranklists of data stream measurements at each acquisition time, as it has been shown that these indices can effectively detect process mean shifts or anomalies in downward and upward directions in a nonparametric manner \cite{Qiu01052001,qiunonpara}. Recently, \citet{quants_paper} proposed a quantile-based nonparametric CUSUM procedure to monitor heterogeneous processes for the first time, thereby significantly expanding the capability of SPC-based monitoring techniques. \citet{JIANG20231623} tested this quantile-based SPC framework on the Tennessee Eastman Process (TEP) benchmark problem \cite{tep_downs_vogel} and showed that it outperformed existing fault detection algorithms in terms of detection speed. \citet{mola} further incorporated this quantile-based CUSUM procedure \cite{quants_paper} in a multi-block orthogonal LSTM autoencoder framework and achieved synergistic improvement in fault detection speed and accuracy in the TEP problem compared to state-of-the-art approaches.

In terms of fault classification, a new research direction propelled by \citet{smith_TDA} and \citet{smith_1} is the use of topological data analysis (TDA) and non-Euclidean geometry (e.g., Riemannian geometry) to more effectively extract the fundamental topological and geometric feature underneath the data streams to be analyzed. Specifically, \citet{smith_1} studied the use of principal geodesic analysis (PGA), a counterpart of PCA applied on the tangent space of the Riemannian manifold, to analyze the TEP dataset and achieved remarkable success in fault classification.

\begin{figure}[t]
    \centering
    \includegraphics[width=\linewidth]{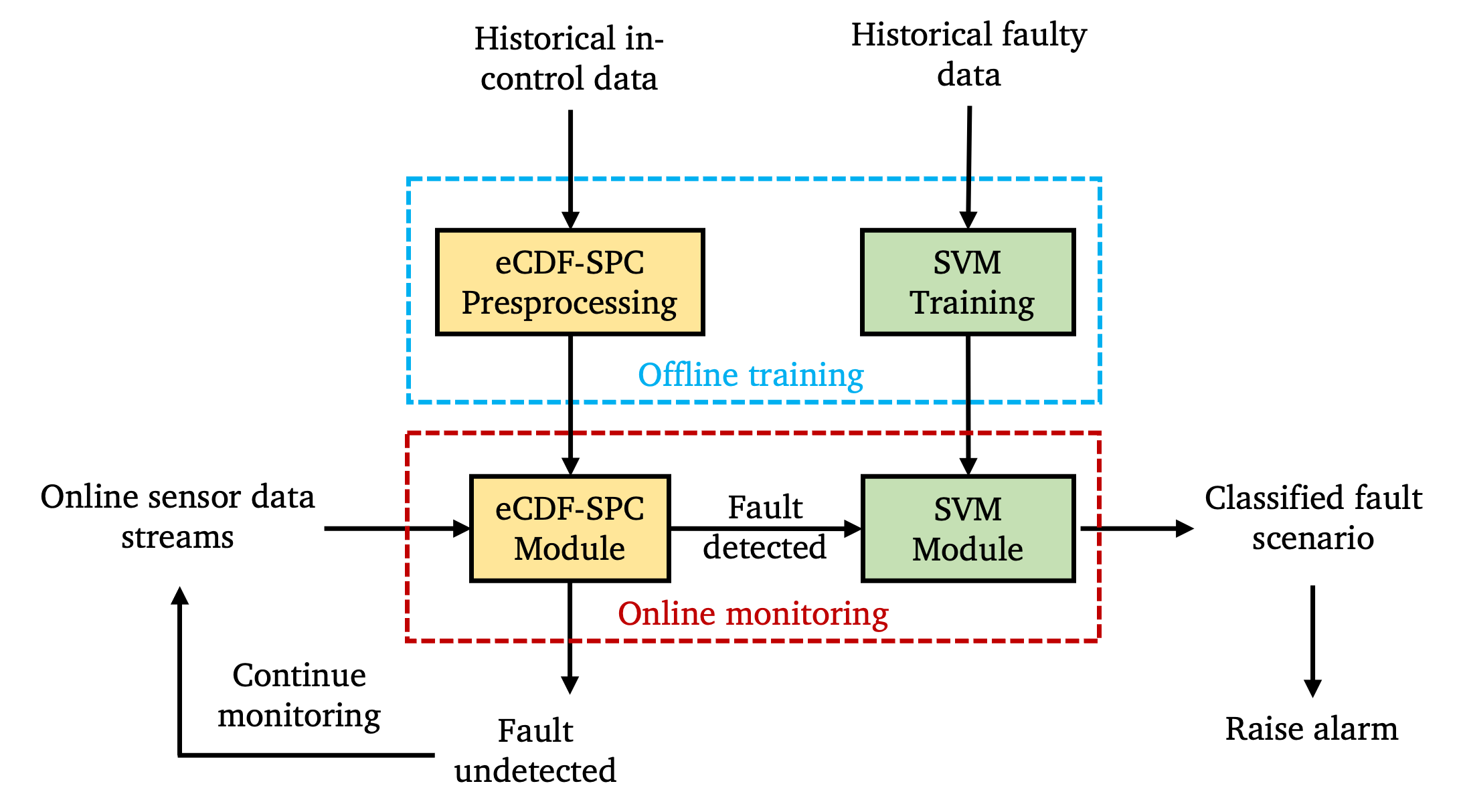}
    \caption{Our FARM framework \cite{alireza:2024:focapd} for holistic fault detection and diagnosis.}
    \label{fig:SAFIRE}
\end{figure}

These exciting breakthroughs in fault detection and classification motivate us to develop a holistic process monitoring framework called FARM, which stands for ``Fast, Accurate and Robust process Monitoring'' in our earlier work \cite{alireza:2024:focapd} originally presented at the FOCAPD 2024 Conference. As shown in Figure \ref{fig:SAFIRE}, FARM has a hierarchical structure that decomposes process monitoring into fault detection task followed by fault classification task. Each task is conducted by the targeted algorithm. Only if a process fault is detected will the online data be sent for classification. Nevertheless, due to page limit requirement, we were unable to conduct a systematic analysis of the FARM monitoring framework. In particular, questions remain on 1) how well FARM performs in handling the intrinsic trade-off between fault detection speed and accuracy and 2) how well FARM could address the somewhat conflicting objectives between fault detection (``the sooner the better'') and fault classification (whose accuracy improves with increasing time due to larger data size). Although they are clearly important, these aspects have historically been overlooked, as most existing fault classification algorithms use the full-size (spanning $\approx 15$ hours) TEP datasets (e.g., \cite{tep_downs_vogel,tepdata,tepdata_gui}) for training and testing, posing the question of how these algorithms would perform in online monitoring setting (i.e., with only part of the dataset being available).

To answer these questions, in this work, we conduct a systematic analysis of the FARM framework on the TEP benchmark problem as a continuation of our FOCAPD 2024 conference proceeding \cite{alireza:2024:focapd}. For fault detection, we adopt the latest advancement in multivariate nonparametric SPC for generalized heterogeneous process monitoring based on empirical cumulative distribution function (eCDF) estimation \cite{ecdf}. For fault classification, we present a modified SVM trained by historical faulty data's covariance matrices that are mapped to the tangent space of a Riemannian manifold. We show that our FARM framework achieves competitive performance compared to state-of-the-art methods, especially in detecting widely-known hard-to-detect faults.

\section{The FARM Monitoring Framework}
\subsection{Fault Detection Module}
As shown in Figure \ref{fig:SAFIRE}, the FARM framework consists of two distinct modules, one for fault detection and the other for fault classification. In fault detection module, we adopt the state-of-the-art eCDF-based SPC algorithm recently developed by \citet{ecdf}. Suppose we are monitoring a total of $p$ data streams simultaneously, whose measurements are denoted as $\mathbf{x}(t) = (x_1(t), \dots, x_p(t))$ over the sampling time $t=1,2,$ and so on. Here, $x_i(t)$ is assumed to be independently and identically distributed (i.i.d.) across time $t$ for every $i = 1,\dots,p$ during the in-control or out-of-control process. This i.i.d. assumption of data streams is often satisfied when $x_i{t}$ are the residual values of data stream measurements \cite{meiiid}, and has been widely used in monitoring industrial big data streams \cite{Zou03072015}. First, we construct the local statistic characterization for univariate data stream $x_i(t)$. Let $\mu_i(t)$ be the CDF value of $x_i(t)$ for sensor $i$ at time $t$. And let $\mathbf{x}_i^0 = (x_{i,1}^0,\dots,x_{i,s_i}^0)$ be a set of historical in-control data from sensor $i$. Now, consider an online measurement $x_i(t)$ and compare it with $\mathbf{x}_i^0$. From the i.i.d. assumption, the prior distribution of $\mu_i(t)$ is $U(0,1)$. Furthermore, by defining an indicator function $y_{i,j}(t) = \mathbf{1}(x_{i,j}^0 < x_i(t))$ for every $j = 1,\dots, s_i$, we can track where $x_i(t)$ stands relative to the historical in-control data collected for sensor $i$. Due to the i.i.d. assumption, the random variable $y_{i,j}(t)$ is also i.i.d. and follows a Bernoulli distribution with respect to the CDF value $\mu_i(t)$:
\begin{equation}\label{eqn_bernoulli}
    \mathbb{P}\left(y_{i,j}(t)|\mu_i(t)\right) \propto \left[\mu_i(t)\right]^{y_{i,j}(t)} \left[1-\mu_i(t)\right]^{1-y_{i,j}(t)}.
\end{equation}
The posterior distribution of $\mu_i(t)$ is given by:
\begin{equation}\label{eqn_posterior}
    \mathbb{P}\left(\mu_i(t)|y_{i,1}(t),y_{i,2}(t),\dots, y_{i,s_i}(t)\right)  \propto \mathbb{P}(\mu_i(t)) \times \prod_{j=1}^{s_i} \mathbb{P}\left(y_{i,j}(t)|\mu_i(t)\right).
\end{equation}
Noting that $\mathbb{P}(\mu_i(t)) = 1$ and substituting Equation \eqref{eqn_bernoulli} into \eqref{eqn_posterior} will give:
\begin{equation*}
    \mathbb{P}\left(\mu_i(t)|y_{i,1}(t),y_{i,2}(t),\dots, y_{i,s_i}(t)\right) \propto [\mu_i(t)]^{\sum_{j=1}^{s_i} y_{i,j}}[1-\mu_i(t)]^{s_i - \sum_{j=1}^{s_i} y_{i,j}(t)},
\end{equation*}
which corresponds to the probability density function of a Beta distribution with parameters $\sum_{j=1}^{s_i} y_{i,j}(t) + 1$ and $s_i - \sum_{j=1}^{s_i} y_{i,j}(t) + 1$. With this, the CDF value at $x_i(t)$, $\mu_i(t)$, can be readily estimated using the posterior mean of $\mathrm{Beta}(\sum_{j=1}^{s_i} y_{i,j}(t) + 1, s_i - \sum_{j=1}^{s_i} y_{i,j}(t) + 1)$ as:
\begin{equation}
    \hat{\mu}_i(t) = \frac{\sum_{j=1}^{s_i} y_{i,j}(t) + 1}{s_i + 2}.
\end{equation}

Essentially, when the data stream is in control, the CDF value $\mu_i(t)$ follows $U(0,1)$, whereas when the data stream has a positive (resp. negative) mean shift, $\mu_i(t)$ will approach toward $1$ (resp. $0$). In other words, the key idea behind the eCDF-based SPC algorithm is to convert the detection of mean shifts in $x_i(t)$ into detecting shifts in the distribution of $\mu_i(t)$ \cite{ecdf}. Nevertheless, as pointed out by \citet{ecdf}, directly using the eCDF $\hat{\mu}_i(t)$ for local statistic characterization can pose some implementation challenges, as $\hat{\mu}_i(t) \in (0,1)$ is not sensitive enough process mean shifts in the distribution of $x_i(t)$. \citet{ecdf} proposed to increase the sensitivity by taking log of the $\hat{\mu}_i(t)$ value, thereby resulting in the following one-sided local statistics $W_i^+(t)$ and $W_i^-(t)$ for detecting positive ($+$) or negative ($-$) mean shift for sensor $i$:
\begin{equation*}
    \begin{aligned}
        & W_i^+(t) = \max\left\{ W_i^+(t-1) - \log{(1-\hat{\mu}_i(t))} - k,0\right\},\\
        & W_i^-(t) = \max\left\{ W_i^-(t-1) - \log{(\hat{\mu}_i(t))} - k,0\right\},\\
        & W_i^+(0) = W_i^-(0) = 0,
    \end{aligned}
    \qquad \forall i=1,\dots,p;\;t=1,2,\dots,
\end{equation*}
where $k>0$ is an allowance parameter, which restarts the CUSUM procedure if no evidence of mean shift is detected after some time. To detect mean shift in either direction, a two-sided local statistic is defined as:
\begin{equation}
    W_{i}(t) = \max\{W_i^+(t), W_i^-(t)\} \qquad \forall i=1,\dots,p;\;t=1,2,\dots.
\end{equation}

Finally, the top-$r$ approach by \citet{mei} is used to obtain the global statistic for monitoring all data streams. The idea is to rank the two-sided local statistic $W_i(t)$ for all $i$ at every sampling time $t$ from the largest to the smallest, and the global statistics is the summation of the highest $r$ local statistics in the ranklist:
\begin{equation}
    V(t) = \sum_{i=1}^{r} W_{(i)}(t),
    \label{eq: global_stat}
\end{equation}
where $(i)$ corresponds to the index of the $i^{\text{th}}$ largest two-sided local statistic. An alarm will be raised when $V(t)$ exceeds a threshold value $H$, which is determined once an in-control average run length ($\mathrm{ARL}_0$) is set by the operator based on the false alarm rate (FAR). A commonly used $\mathrm{ARL}_0$ is 370, which corresponds to a false alarm rate of no more than $0.0027$ (the $3\sigma$ limit widely used in quality control \cite{montgomery2008}). Therefore, this top-r-based stopping criterion not only has concrete theoretical foundation and statistical significance, but also offers great flexibility as operators can customize the choice of $H$ based on the relative severity and tolerance of process failures. An in-depth discussion on the theory and application of this eCDF-based CUSUM procedure can be found in \citet{ecdf}.

\subsection{Fault Classification Module} \label{sec: diagnosis}

Once an alarm is raised during online monitoring, the data streams will start to be sent to a modified SVM-based fault classification module that utilizes the covariance matrix derived from $\mathbf{x}(t)$. By this time, although an alarm is raised and operators are alerted, there are typically not enough faulty data to exactly pinpoint which fault the process has. Therefore, to improve the classification accuracy, we introduce a tunable hyperparameter called ``patience time'' $t_p$, which corresponds to the number of additional sampling times after the alarm is first raised until the fault is classified. As expected, $t_p$ must be long enough to provide meaningful insights for fault classification module to give accurate prediction, but should not be too long such that the process monitoring framework becomes useless. Once the patience time elapses, FARM would automatically select the process data from the last $t_w$ sampling times called the ``window length'' for classification. This is not only because the fault classification module can only take a fixed number of samples to construct the covariance matrix, but also to reduce computational costs and mitigate the need to transfer a large amount of data.

To train the SVM model using the historical sensor data corresponding to different fault scenarios, we first compute the covariance matrix of the historical faulty data streams, followed by training the SVM model over the covariance matrix instead of the original faulty data streams. This modification is inspired by the fact that covariance matrices are symmetric and positive definite, and thus always lie on a Riemannian manifold. It has been recently shown that, by respecting this geometric insight, one can greatly enhance the accuracy and interpretability of classification, regression, dimensionality reduction algorithms by conducting these computations on the tangent space of the manifold \cite{smith_1}. Inspired by this result, we map the generated covariance matrices to their tangent space, which intersect the Riemannian manifold where these covariance matrices reside at the geometric mean of the covariance matrices. Through the logarithmic map of Equation \eqref{eq: mapping}, the set of covariance matrices of sensor data streams $\mathbf{C}_m$, which lie on a Riemannian manifold, will be projected onto a tangent space with minimal geometric distortion:
\begin{equation}
    \hat{\mathbf{C}}_m = \log_{\bar{\mathbf{C}}} (\mathbf{C}_m)
    \label{eq: mapping}
\end{equation}
where $\mathbf{C}_m, \, \bar{\mathbf{C}}$ and $\hat{\mathbf{C}}_m \in \mathbb{R}^{p \times p}$ are covariance matrix of sensor data streams for dataset $m$, the geometric mean of the set of matrices $\mathbf{C}_m$ and the mapped covariance matrix for dataset $m$, respectively. After this data preprocessing step, the mapped covariance matrices are used as input features, whereas the corresponding fault scenarios are used as labels to train a standard SVM model using a radial basis function (RBF) kernel.


\subsection{Overall Workflow}
Here, we summarize the overall workflow for conducting offline training and online monitoring in FARM monitoring framework:
\begin{itemize}
    \item Offline Training:
    \begin{enumerate}
        \item Historical in-control data of the plant are gathered, normalized and standardized. At this step the means ($\boldsymbol{\mu}$) and standard deviations ($\boldsymbol{\sigma}$) of the process variables are stored for later use. 
        \item For a given set of parameters $k$, $r$, and $\mathrm{ARL}_0$, eCDF-based SPC module is trained to obtain the control limit $H$, which corresponds to the pre-specified $\mathrm{ARL}_0$.
        \item Historical faulty data of the plant are gathered, normalized and standardized using  in-control $\boldsymbol{\mu}$ and $\boldsymbol{\sigma}$ obtained at Step 1. 
        \item Normalized faulty data are monitored by the eCDF-based SPC. If the global statistics $V(t) \geq H$ at time $t$, an alarm is raised.
        
        \item Monitored faulty data are gathered and a patience time $t_p$ is chosen as number of samples to gather after the alarm has raised. The choice of $t_p$ depends on the required accuracy. 

        \item The obtained time series are then preprocessed as only the last time steps with the size of window length $t_w$ is chosen for further analysis. These time series are called ``sub-time series'' as they are only a subset of the gathered \st{monitored} time series.
        
        \item The covariance matrix $\mathbf{C}_m$ of sensor data streams for each sub-time series $m$ is calculated.
        
        \item The covariance matrices are mapped $\mathbf{C}_m$ from the Riemannian manifold to the tangent space to obtain the mapped covariance matrices $\hat{\mathbf{C}}_m$. The mean $\mathbf{\Bar{C}}$ is also stored at this stage for online monitoring.
        
        \item The mapped covariance matrices $\hat{\mathbf{C}}_m$ are flattened and used as the features for training the SVM model with the corresponding fault number as the labels.
    \end{enumerate}
    
    \item Online Monitoring:
    \begin{enumerate}
        \item Online sensor data $\mathbf{x}(t)$ are normalized and standardized using the historical in-control $\boldsymbol{\mu}$ and $\boldsymbol{\sigma}$ obtained during offline training.
        
        \item Normalized online data are monitored by the trained eCDF-SPC module. At time $t$, if $V(t) \geq H$, an alarm is raised. If not, eCDF-SPC algorithm continues to monitor the process variables.
        
        \item If a fault is detected, the eCDF-SPC algorithm will wait for the set patience time $t_p$ from offline training to gather more faulty data.
        
        \item The last $t_w$ samples are selected prior to covariance calculation.
        
        \item The covariance matrix $\mathbf{C}$ of the selected samples is calculated.
        
        \item The covariance matrix $\mathbf{C}$ is mapped using the obtained mean $\mathbf{\Bar{C}}$ from offline training to obtain the mapped covariance matrix $\hat{\mathbf{C}}$.
        
        \item The mapped covariance matrix $\hat{\mathbf{C}}$ is flattened and fed to the trained SVM model in the offline training to obtain the predicted fault label.
    \end{enumerate}
    
\end{itemize}

\section{Case Study: Tennessee Eastman Process}

\begin{figure}[t!]
    \centering
    \includegraphics[width=\textwidth]{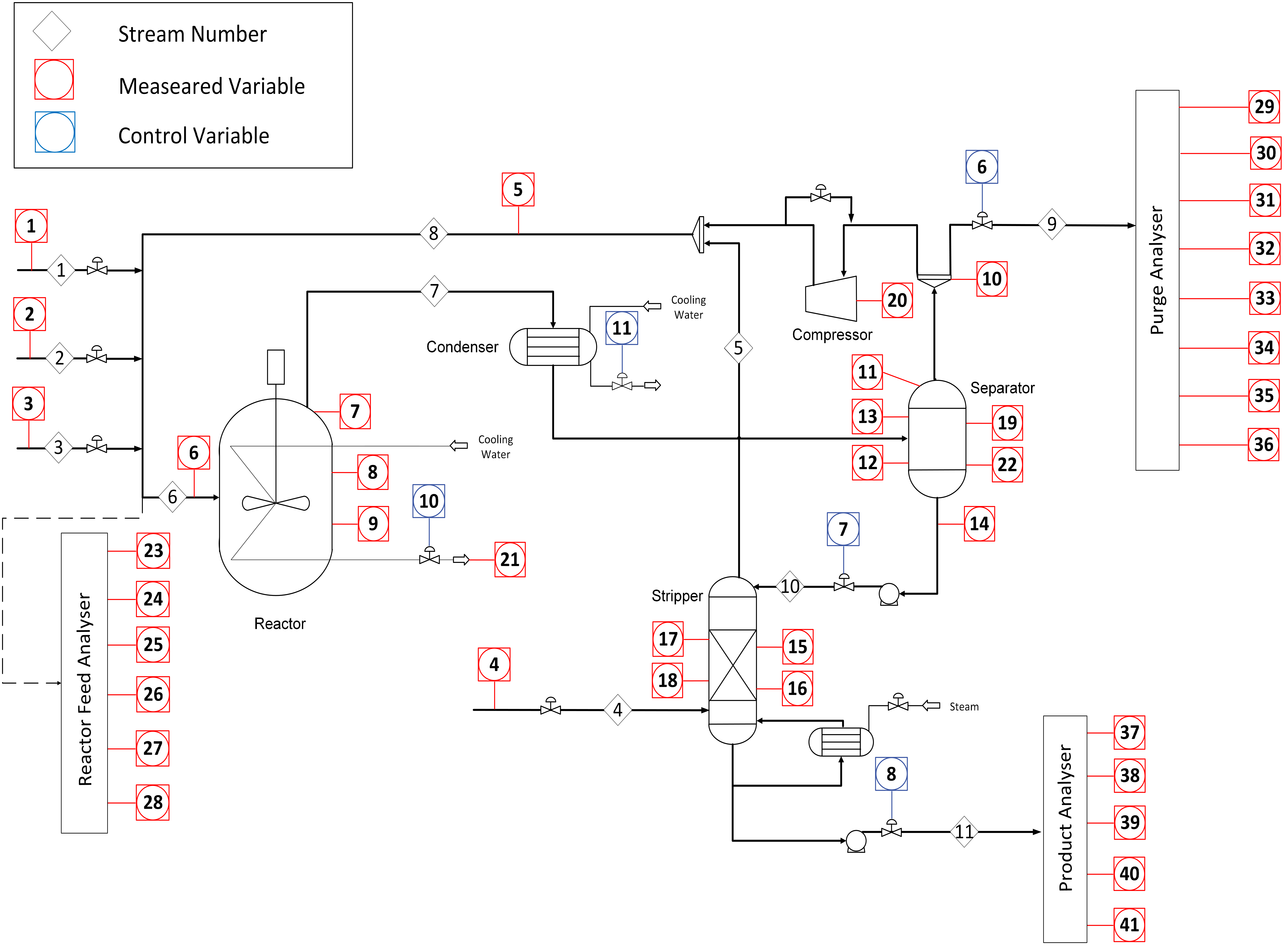}
    \caption{Schematic of Tennessee Eastman Process \cite{tep_downs_vogel}.}
    \label{fig: tep_pfd}
\end{figure}

The Tennessee Eastman Process (TEP) is a simulation process developed by the Eastman Chemical Company based on an actual chemical process \cite{tep_downs_vogel}. The TEP has been widely adopted as a benchmark for chemical process control, optimization, and monitoring. As illustrated in the process flow diagram of Figure \ref{fig: tep_pfd}, the TEP contains five major unit operations, which are associated with 12 manipulated variables and 41 measured variables in total. Table \ref{table_TEPfaults} shows the description of each of the 20 faults considered in the TEP problem. For this case study, we synthesize TEP dataset for this case study using the MATLAB/Simulink-based Graphical User Interface (GUI) \cite{tepdata_gui}. Table \ref{table_TEPvariables} lists the variables considered in this study. Six variables are excluded from our analysis since they remain constant in the dataset.

In this study, we evaluate the performance of our FARM framework against four benchmark process monitoring algorithms that we build in house, namely PCA-$T^2$, Quantile-based SPC (Q-SPC) \cite{JIANG20231623,quants_paper}, Two Class Nonlinear Support Vector Machine (TCNL-SVM) \cite{svm_1}, and Long Short-Term Memory Auto Encoder (LSTM AE) \cite{REN2020102}. These selections cover a spectrum of existing approaches, ranging from dimensionality reduction-based methods, to machine learning-based methods, to SPC-based methods. Three performance metrics, namely fault detection rate (FDR), fault detection speed (FDS), and false alarm rate (FAR), are used to evaluate and compare the performance of different algorithms. The FDR quantifies the model's ability to correctly detect faults when they occur. It is defined as the ratio of the number of correctly identified faulty samples ($N_{\mathrm{TP}}$) to the total number of actual faulty samples ($N_{\mathrm{TP}} + N_{\mathrm{FN}}$). A high FDR indicates that the framework is effective in detecting faults without missing significant events. The FDS, defined as the number of additional sampling times required to detect a fault after it occurs, measures the responsiveness of the model. A lower FDS reflects faster detection. Lastly, the FAR evaluates the framework's ability to minimize false positives, where normal operating conditions are incorrectly flagged as faults. It is defined as the ratio of the number of normal samples misclassified as faulty ($N_{\mathrm{FP}}$) to the total number of normal samples ($N_{\mathrm{TN}} + N_{\mathrm{FP}}$). A low FAR is critical to ensuring the reliability of the monitoring system and avoiding unnecessary interventions during normal operations. Together, these metrics collectively provide a comprehensive assessment of process monitoring algorithm performance.

\begin{table}[ht!]
\centering
\caption{The detailed information of the monitoring variables used in our case study.}\label{table_TEPvariables}
\begin{adjustbox}{width=\columnwidth}
    \begin{tabular}{ccc|ccc}
    \toprule
    \textbf{No.} & \textbf{Variable} & \textbf{Description}         & \textbf{No.} & \textbf{Variable} & \textbf{Description}                       \\ \midrule
    \textbf{1}   & XMEAS1 & A feed                                  & \textbf{24}          & XMEAS24              & Reactor Feed Composition of B            \\
    \textbf{2}   & XMEAS2  & D feed                                 & \textbf{25}          & XMEAS25            & Reactor Feed Composition of C                       \\
    \textbf{3}   & XMEAS3  & E feed                                 & \textbf{26}          & XMEAS26              & Reactor Feed Composition of D                        \\
    \textbf{4}   & XMEAS4  & A and C feed                           & \textbf{27}          & XMEAS27             & Reactor Feed Composition of E                            \\
    \textbf{5}   & XMEAS5  & Recycle flow                           & \textbf{28}          & XMEAS28             & Reactor Feed Composition of F   \\
    \textbf{6}   & XMEAS6  & Reactor feed rate                      & \textbf{29}          & XMEAS29            & Purge composition of A \\
    \textbf{7}   & XMEAS7  & Reactor pressure                        & \textbf{30}          & XMEAS30             & Purge composition of B                     \\
    \textbf{8}   & XMEAS8  & Reactor level                           & \textbf{31}          & XMEAS31            & Purge composition of C                     \\
    \textbf{9}   & XMEAS9  & Reactor temperature                     & \textbf{32}          & XMEAS32             & Purge composition of D                      \\
    \textbf{10}   & XMEAS10  & Purge rate                               & \textbf{33}          & XMEAS33             & Purge composition of E               \\
    \textbf{11}  & XMEAS11  & Product separator temperature           & \textbf{34}          & XMEAS34              & Purge composition of F                   \\
    \textbf{12}  & XMEAS12  & Product separator level                 & \textbf{35}          & XMEAS35             & Purge composition of G       \\
    \textbf{13}  & XMEAS13  & Product separator pressure              & \textbf{36}          & XMEAS36             & Purge composition of H        \\
    \textbf{14}  & XMEAS14  & Product separator underflow               & \textbf{37}          & XMEAS37             & Product composition of D                  \\
    \textbf{15}  & XMEAS15  & Stripper level                          & \textbf{38}          & XMEAS38             & Product composition of E                \\
    \textbf{16}  & XMEAS16  & Stripper pressure                      & \textbf{39}          & XMEAS39             & Product composition of F                \\
    \textbf{17}  & XMEAS17  & Stripper underflow                         & \textbf{40}          & XMEAS40             & Product composition of G               \\
    \textbf{18}  & XMEAS18  & Stripper temperature                          & \textbf{41}          & XMEAS41             & Product composition of H               \\
    \textbf{19}  & XMEAS19  & Separator steam flow                          & \textbf{42}          & XMV6             & Purge valve               \\
    \textbf{20}  & XMEAS20  & Compressor work                          & \textbf{43}          & XMV7            & Separator pot liquid flow              \\
    \textbf{21}  & XMEAS21  & Reactor cooling water \\ & & outlet temperature                         & \textbf{44}          & XMV8            & Stripper liquid product flow               \\
    \textbf{22}  & XMEAS22  & Separator cooling water \\ & & outlet temperature                          & \textbf{45}          & XMV10             & Reactor cooling water valve               \\
    \textbf{23}  & XMEAS23  & Reactor Feed Composition of A                          & \textbf{46}          & XMV11             & Condenser cooling water flow              
     \\ \bottomrule    
    \end{tabular}
\end{adjustbox}
\end{table}

\begin{table}[ht!]
\centering
\caption{The detailed information of TEP faults. \label{table_TEPfaults}}
    \begin{tabular}{ccc}
    \toprule
    \textbf{Fault No.} & \textbf{Process variable}                              & \textbf{Type}    \\ \midrule
    \textbf{1}         & A/C feed ratio, B composition constant (stream 4)      & Step             \\
    \textbf{2}         & B composition, A/C ratio constant (stream 4)           & Step             \\
    \textbf{3}         & D feed temperature (stream 2)                          & Step             \\
    \textbf{4}         & Reactor cooling water inlet temperature                & Step             \\
    \textbf{5}         & Condenser cooling water inlet temperature              & Step             \\
    \textbf{6}         & A feed loss (stream 1)                                 & Step             \\
    \textbf{7}         & C header pressure loss-reduced availability (stream 4) & Step             \\
    \textbf{8}         & A, B, C feed composition (stream 4)                    & Random variation \\
    \textbf{9}         & D feed temperature (stream 2)                          & Random variation \\
    \textbf{10}        & C feed temperature (stream 4)                          & Random variation \\
    \textbf{11}        & Reactor cooling water inlet temperature                & Random variation \\
    \textbf{12}        & Condenser cooling water inlet temperature              & Random variation \\
    \textbf{13}        & Reaction kinetics                                      & Slow drift       \\
    \textbf{14}        & Reactor cooling water valve                            & Sticking         \\
    \textbf{15}        & Condenser cooling water valve                          & Sticking         \\
    \textbf{16}        & Variation coefficient of steam supply                  & Unknown          \\
    \textbf{17}        & Variation coefficient of reactor heat transfer         & Unknown          \\
    \textbf{18}        & Variation coefficient of condenser heat transfer       & Unknown          \\
    \textbf{19}        & Unknown                                                & Unknown          \\
    \textbf{20}        & Unknown                                                & Unknown          \\ \bottomrule
    \end{tabular}
\end{table}

\subsection{Fault Detection Performance}

To evaluate the fault detection performance, $12000$ in-control simulations, each containing $1400$ sampling times (sampling frequency is around $10$ seconds), are generated using the MATLAB/Simulink GUI \cite{tepdata_gui}. These in-control simulations are then split into $9600$ and $2400$ simulations for training and testing, respectively. The threshold value $H$ used in the SPC algorithm is obtained from the $9600$ in-control simulations by setting the in-control average run length ($\mathrm{ARL}_0$) to be $500$ with $r=4$ and an optimal allowance parameter of $k=1.3$. An $\mathrm{ARL}_0$ value of 500 means that, on average, FARM raises $1$ false alarm every $500$ sampling times when the process is in control (i.e., the average $\mathrm{FAR} = \frac{1}{500} = 0.002$). Under these parameter settings, the threshold value $H$ is determined to be $1546.875$. A similar procedure is followed to determine the threshold value corresponding to $\mathrm{ARL}_0 = 500$ for Q-SPC. Table \ref{tab:FAR} lists the $\mathrm{FAR}$ determined from the $2400$ simulations in the test set. Both eCDF-based SPC and Q-SPC methods report $\mathrm{FAR}$ values that are very close to the originally set value ($0.02$) in the training set, indicating that the calculated $H$ in the training set is transferrable to testing set. Also, this result highlights the flexibility of SPC-based fault detection algorithms in selecting desired $\mathrm{FAR}$ based on operator's preference. On the other hand, TCNL-SVM and LSTM-AE have no control over $\mathrm{FAR}$ and the $\mathrm{FAR}$ results cannot be transferred between training set and test set, as well as between offline training and online monitoring.

\begin{table}[h]
    \caption{Comparison of studied methods in terms of average FAR over 2400 in-control simulations}
    \label{tab:FAR}
    \centering
    \begin{tabular}{c c c c c c}
        \toprule
         Model &  eCDF-SPC &  Q-SPC & PCA-$T^2$ & TCNL-SVM & LSTM AE \\
         \midrule
         FAR& \textbf{0.0021} & 0.0025& 0.0036& 0.1620 &0.0086\\
         \bottomrule
    \end{tabular}
\end{table}
 
Next, we conduct $500$ out-of-control simulations for each fault with a duration of $\approx 20$ hours in the simulation environment (which corresponds to 7000 sampling steps). In Table \ref{tab: FDR_compare}, we summarize the average $\mathrm{FDR}$ over $500$ out-of-control simulations, from which one can see that our eCDF-based SPC algorithm outperforms benchmark methods in most faults. In particular, it is worth mentioning that eCDF-based SPC algorithm achieves remarkable $\mathrm{FDR}$ for detecting Faults 3, 9 and 15, which are incipient faults that are known to be challenging to detect \cite{Agarwal2022HierarchicalProcess}. In fact, Q-SPC also performs reasonably well in detecting these three faults, compared to the other methods. The $\mathrm{FDR}$ results suggest that SPC algorithms are better at keeping the alarm raised as long as a fault occurs (for theoretical justifications, see Properties 1 and 2 in \citet{ecdf}). 

\begin{table}[!ht]
    \centering
    \caption{Comparison of the methods in terms of average FDR over 500 out-of-control simuatlions}
    \label{tab: FDR_compare}
    \begin{tabular}{c c c c c c}
    \toprule
         Fault &  eCDF-SPC &  Q-SPC & PCA-$T^2$ & TCNL-SVM & LSTM AE \\
         number & & & &  & \\
         \midrule
         1 & 0.9505& 0.9422& 0.9808&  0.9774& \textbf{0.9823}\\
         2 & \textbf{0.9811}& 0.9802& 0.9761 & 0.9724 & 0.9775\\
         3 & \textbf{0.9708} & 0.9462&  0.0282 & 0.0001& 0.0084\\
         4 & 0.9590& 0.9096& 0.9997& 0.8840 & \textbf{1.0000}\\
         5 & \textbf{0.9896}& 0.9872& 0.1519 & 0.1719& 0.0730\\
         6 & 0.9895& 0.9765& \textbf{1.0000} & \textbf{1.0000} & \textbf{1.0000}\\
         7 & 0.9693& 0.9449& \textbf{1.0000} & 0.9934& \textbf{1.0000}\\
         8 & \textbf{0.9687}& 0.9616& 0.9528 & 0.9610& 0.9478\\
         9 & \textbf{0.9630}& 0.8277& 0.0197 & 0.2923& 0.0078\\
         10 & \textbf{0.9648}& 0.9418& 0.5158 & 0.7559& 0.6962\\
         11 & \textbf{0.9808}& 0.9772& 0.9868 & 0.8228& 0.9861\\
         12 & \textbf{0.9763}& 0.9701& 0.9601 & 0.9475& 0.9482\\
         13 & \textbf{0.9643}& 0.9267& 0.8932 & 0.7572& 0.8932\\
         14 & 0.9840 & 0.9688& 0.9864 &  0.9324& \textbf{0.9878}\\
         15 & \textbf{0.9599}& 0.4841& 0.0115 & 0.4638& 0.0072\\
         16 & \textbf{0.9729}& 0.9625& 0.8317 & 0.8724& 0.8157\\
         17 & \textbf{0.9648}& 0.9301& 0.9060 & 0.9315& 0.9022\\
         18 & \textbf{0.9621}& 0.8922& 0.7052 & 0.7319& 0.7798\\
         19 & \textbf{0.9606}& 0.4987& 0.0120 & 0.2746& 0.0070\\
         20 & \textbf{0.9617}& 0.9186& 0.8441 & 0.9342& 0.8730\\
         \midrule
         Average & \textbf{0.9697} & 0.8973 & 0.6881 & 0.7338 & 0.6947 \\
         \bottomrule
    \end{tabular}
\end{table}

\begin{figure}[!ht]
    \centering
    \includegraphics[width=0.6\textwidth]{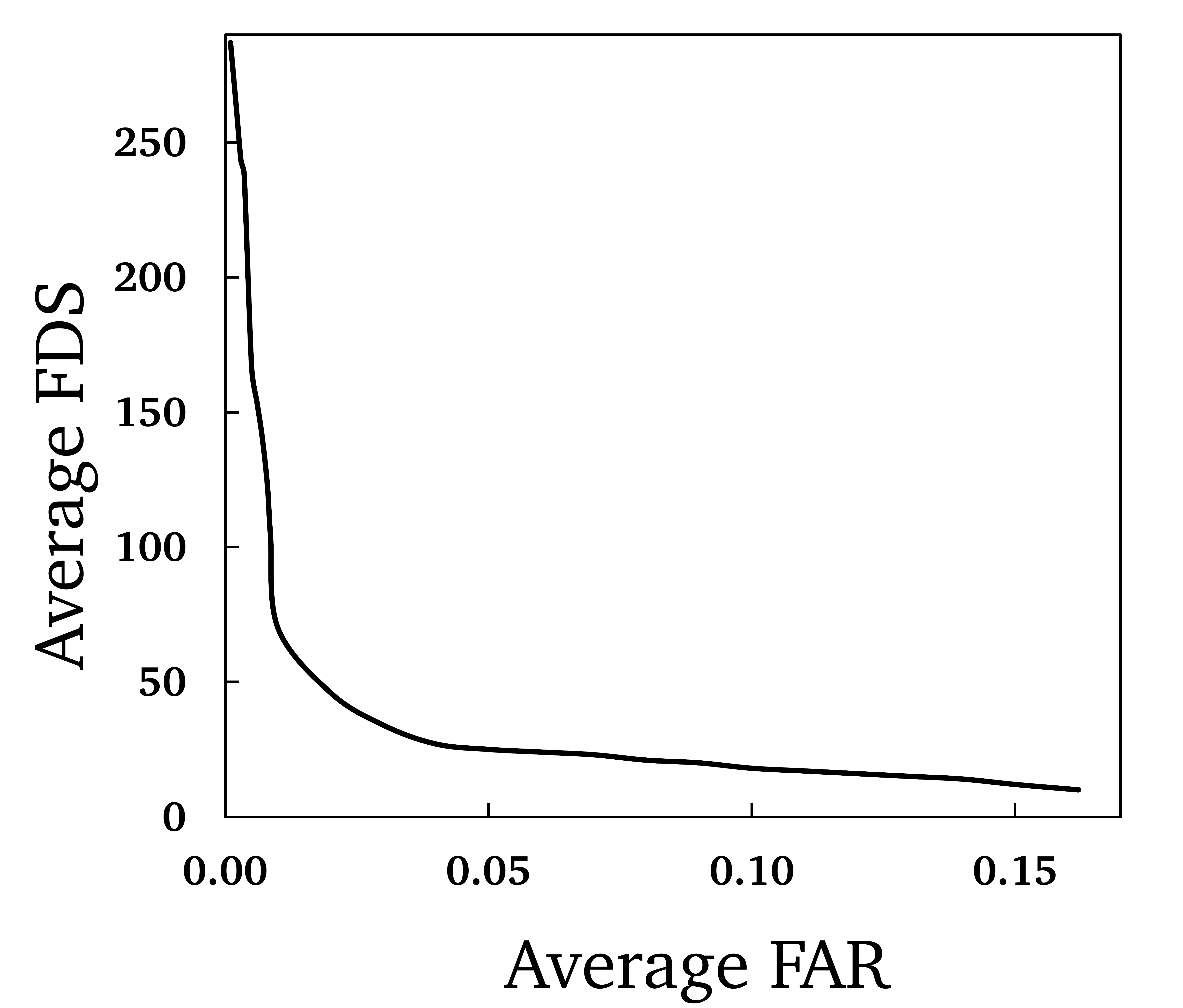}
    \caption{Average FDS versus average FAR for detecting Fault 15 using eCDF-based SPC method. }
    \label{fig: FDD_vs_FAR}
\end{figure}

Meanwhile, Table \ref{tab:FDD_compare} summarizes the $\mathrm{FDS}$ results for different methods under their respective $\mathrm{FAR}$s listed in Table \ref{tab:FAR}. Although SPC-based methods do not perform competitively against other methods, it should be noted that other methods have much higher $\mathrm{FAR}$ values. To understand the trade-off between $\mathrm{FAR}$ and $\mathrm{FDS}$, we leverage the flexibility of SPC-based methods in specifying any desired $\mathrm{FAR}$ and plot the average $\mathrm{FDS}$ against the average $\mathrm{FAR}$ for eCDF-SPC method in detecting Fault 15. It is clear from Figure \ref{fig: FDD_vs_FAR} that the fault detection speed drastically improves as $\mathrm{FAR}$ is relaxed (i.e., increases). 

To ensure a fair comparison, we rerun the eCDF-SPC algorithm by specifying the same $\mathrm{FAR}$ listed in Table \ref{tab:FAR} for the three non-SPC methods. From Table \ref{tab:FDD_compare_similarFAR}, it is clear that, under the same $\mathrm{FAR}$, the eCDF-based SPC algorithm outperforms non-SPC methods in terms of $\mathrm{FDS}$ in many instances, especially at higher $\mathrm{FAR}$.

\begin{table}[!ht]
    \centering
    \caption{Comparison of the methods in terms of average FDS over 500 out-of-control simulations.}
    
    \label{tab:FDD_compare}
    \begin{tabular}{c c c c c c}
    \toprule
         Fault &  eCDF-SPC &  Q-SPC & PCA-$T^2$ & TCNL-SVM & LSTM AE \\
         number & & & & & \\
         \midrule
         1 & 82 & 96 & 27 & 37 & 29\\
         2 & 127 & 124 & 85 & 87 & 112\\
         3 & 202 & 159 &  195 & 34 & 289\\
         4 & 69 & 91 & 0 & 1 & 0\\
         5 & 73 & 90 & 16 & 13 & 30\\
         6 & 69 & 89 & 0 & 0 & 0\\
         7 & 65 & 88 & 0 & 3 & 0\\
         8 & 215 & 217 & 167 & 70 & 213\\
         9 & 251 & 286 & 268 & 98 & 300\\
         10 & 240 & 266 & 222 & 67 & 270\\
         11 & 136 & 147 & 75 & 143 & 88\\
         12 & 166 & 166 & 113 & 87 & 143\\
         13 & 236 & 269 & 216 & 171 & 271\\
         14 & 114 & 128 & 51 & 89 & 72\\
         15 & 263 & 337 & 285 & 53 & 301\\
         16 & 192 & 204 & 140 & 81 & 189\\
         17 & 237 & 272 & 219 & 95 & 246\\
         18 & 234 & 274 & 253 & 100 & 296\\
         19 & 257 & 346 & 269 & 24 & 321\\
         20 & 255 & 277 & 258 & 45 & 283\\
         \midrule
         Average & 174.6 & 196.3 & 142.95 & 64.9 & 172.65\\
         \bottomrule
    \end{tabular}
\end{table}

\begin{table}[!ht]
\centering
\caption{Comparison of average $\mathrm{FDS}$ between eCDF-based SPC algorithm and non-SPC methods under the same $\mathrm{FAR}$.}
\label{tab:FDD_compare_similarFAR}
\begin{adjustbox}{width=\columnwidth}
    \begin{tabular}{c c c | c c | c c}

\toprule

\multirow{2}{*}{Fault} & \multicolumn{2}{c }{$\mathrm{FAR} = 0.0036$} & \multicolumn{2}{c}{$\mathrm{FAR} = 0.1620$} & \multicolumn{2}{c}{$\mathrm{FAR} = 0.0086$} \\ 
\cline{2-7} 
                    
& eCDF-SPC     & PCA-$T^2$   & eCDF-SPC     & TCNL-SVM    & eCDF-SPC     & LSTM AE    \\ 

\midrule
1 & 66 & \textbf{27} & \textbf{6} & 37   & 45  & 29 \\
2 & 106  & \textbf{85} & \textbf{6}  &  87  &  \textbf{75} &  112\\
3 & \textbf{163} & 195 & \textbf{8}  &  34  & \textbf{87}  &  289\\
4 & 49 & \textbf{0}   & 3  &  \textbf{1}  &  33  &  \textbf{0}\\
5 & 51 & \textbf{16}  & \textbf{5}  &  13  & 34  &  \textbf{30}\\
6 & 50 & \textbf{0}   & 4 &  \textbf{0}  &  32 &  \textbf{0}\\
7 & 39 & \textbf{0}   &  \textbf{2}  &  3 &  23 &  \textbf{0}\\
8 & \textbf{142} & 167 & \textbf{6}  &  70 &  \textbf{110} &  213\\
9 & \textbf{228} & 268 &  \textbf{7} &  98  &  \textbf{109} &  300\\
 10 & 226 & \textbf{222} & \textbf{6} &  67   &  \textbf{109} &  270\\
11 & 127 & \textbf{75} & \textbf{9} & 143  & \textbf{85} &  88\\
12 & 149 & \textbf{113} & \textbf{7} &  87 & \textbf{97} &  143\\
13 & \textbf{215} & 216 & \textbf{12} &  171 & \textbf{106} &  271\\
14 & 109 & \textbf{51} & \textbf{8} & 89 & \textbf{71} &  72\\
15 & 238 & \textbf{285} & \textbf{10} & 53 & \textbf{113} &  301\\
16 & 186 & \textbf{140} & \textbf{8} & 81 & \textbf{100}  &  189\\
17 & \textbf{192} & 219 & \textbf{6} & 95 & \textbf{109} &  246\\
18 & \textbf{223} & 253 & \textbf{5} & 100 & \textbf{105} &  296\\
19 & \textbf{237} & 269 & \textbf{7} & 24 & \textbf{108} &  321\\
20 & \textbf{230} & 258 & \textbf{10} & 45 & \textbf{109} &  283 \\
\midrule
Average &  151.3 & 142.95 & 6.75 & 64.9 & 83 & 172.65 \\
\bottomrule

\end{tabular}
\end{adjustbox}
\end{table}

\subsection{Fault Classification Performance}
To prepare the dataset for fault classification, we introduce each fault starting at the $501^{\text{st}}$ sampling time step. Overall, $500$ simulations, each containing $500$ in-control sampling time steps and $7000$ out-of-control sampling time steps, are carried out for each fault, thereby resulting in $10000$ simulations in total. 20\% of the simulations are used for testing. And each training simulation is monitored by the eCDF-SPC algorithm. As discussed in Section \ref{sec: diagnosis}, as soon as a fault is detected by the eCDF-SPC method, our FARM monitoring framework will wait for $t_p$ sampling time steps to acquire more faulty data. When the patience time is reached, FARM will select the datasets acquired within the last $t_w$ sampling time steps for each simulation to calculate the covariance matrix. Specifically, $t_w$ is set to be the average number of sampling time steps from when the fault is introduced to when classification takes place for all training simulations. By specifying a window length $t_w$, we mimic process monitoring in real world, where the exact time when a fault occurs is not known to us. Each of the covariance matrices will be mapped to the tangent space using Equation \eqref{eq: mapping}. Finally, $\hat{\mathbf{C}}_m$ are used as input features to train a SVM model. The model implementation is done in \texttt{sklearn} \cite{scikit-learn} and hyperparameter tuning is conducted using the grid search method.

\begin{figure}[t]
    \centering
    \includegraphics[width=0.6\linewidth]{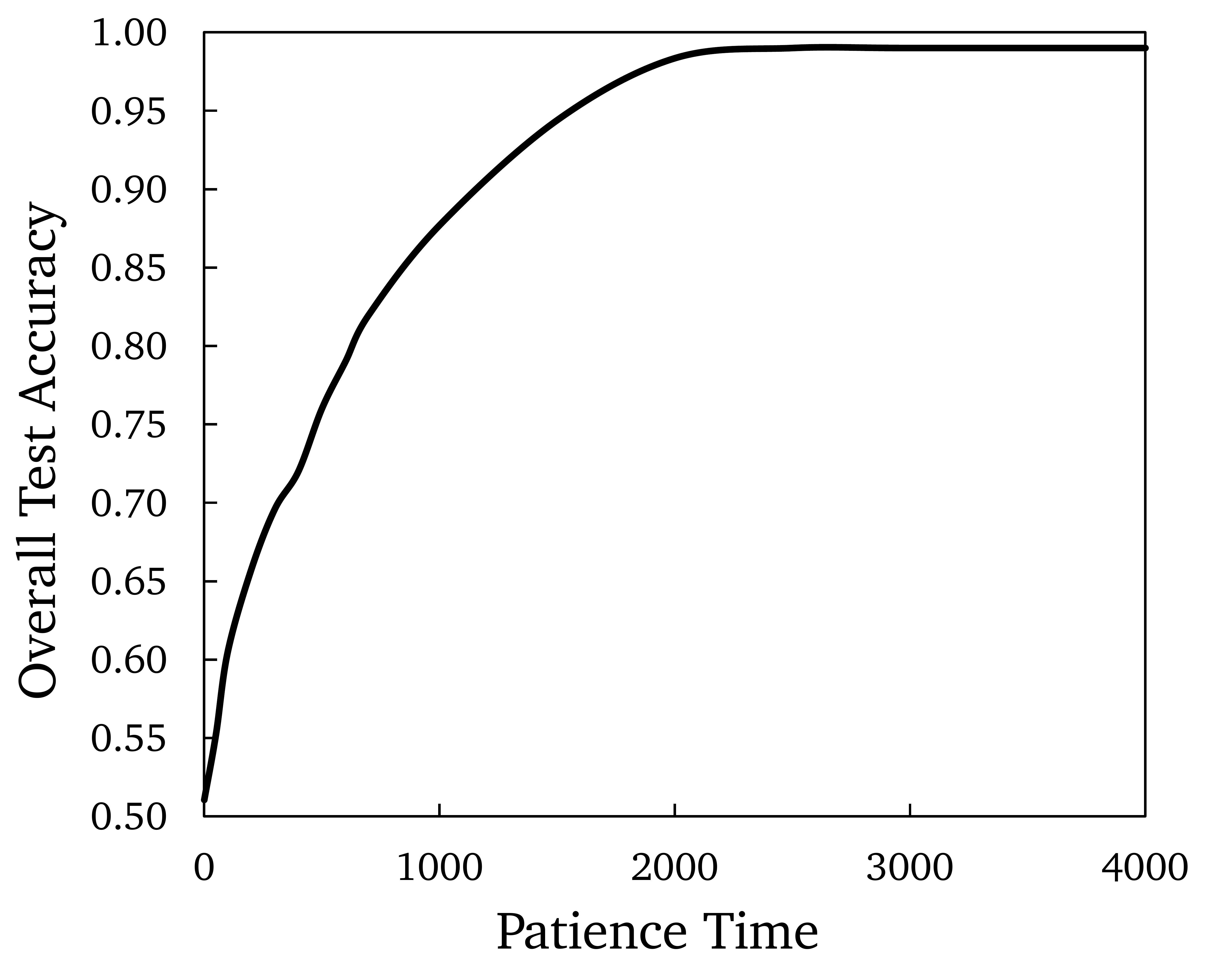}
    \caption{Average fault classification accuracy in the test set for all 20 faults with respect to patience time.} 
    \label{fig: sensitivity_tp}
\end{figure}

To understand how $t_p$ affects fault classification accuracy, we conduct a sensitivity analysis by varying Figure \ref{fig: sensitivity_tp} shows the sensitivity analysis performed by varying the patience time from $100$ to $4000$ sampling time steps and calculating the overall classification accuracy on the test set for all 20 faults. Figure \ref{fig: sensitivity_tp}, we see that: 1) even if $t_p$ is set to be 0 (i.e., we perform fault classification immediately after an alarm is raised), the overall fault classification accuracy is still more than 50\%, 2) the overall fault classification accuracy gradually improves as more patience time is allowed, and 3) there is a diminishing return on the overall classification accuracy when $t_p > 2000$. Based on these results, we set the patience time be $700$ ($\approx 1.9$ hours in simulation environment) for all simulations. And the window length corresponding to $t_p = 700$ is calculated to be $t_w = 873$.

\begin{figure}[!ht]
    \centering
    \includegraphics[width=\textwidth]{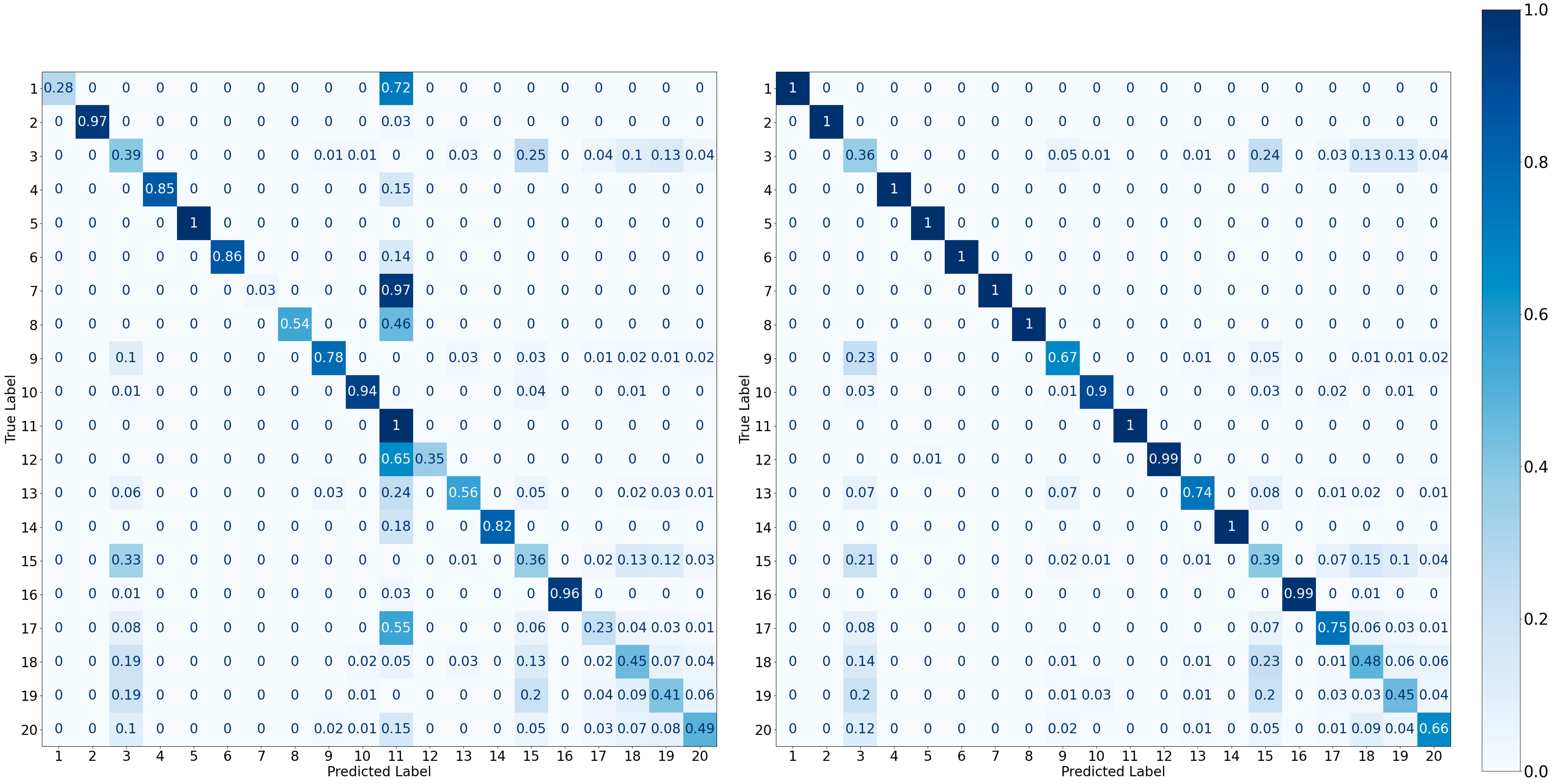}
    \caption{Mapping the covariance matrices to the tangent space improves classification accuracy. (Left): Confusion matrix of SVM method trained on unmapped covariance matrices shows an average fault classification accuracy of 61\%; (Right): Confusion matrix of SVM method trained on mapped covariance matrices shows an improved classification accuracy of 82 \%. The patience time is set to 700 samples for both cases.}
    \label{fig: compare_cm}
\end{figure}

Figure \ref{fig: compare_cm} illustrates how Riemannian geometric analysis improves fault classification accuracy. By introducing a data preprocessing step that maps all covariance matrices to a Riemannian manifold, the overall fault classification accuracy is significantly improved from 61\% to 82\%. For certain faults, such as Fault 19, the relative improvement is greater than 100\%. Furthermore, when comparing the confusion matrix of our modified SVM method with a state-of-the-art supervised LSTM AE method \cite{Agarwal2022HierarchicalProcess}, we see that our modified SVM also performs competitively.

\begin{figure}[!ht]
    \centering
    \includegraphics[width=\textwidth]{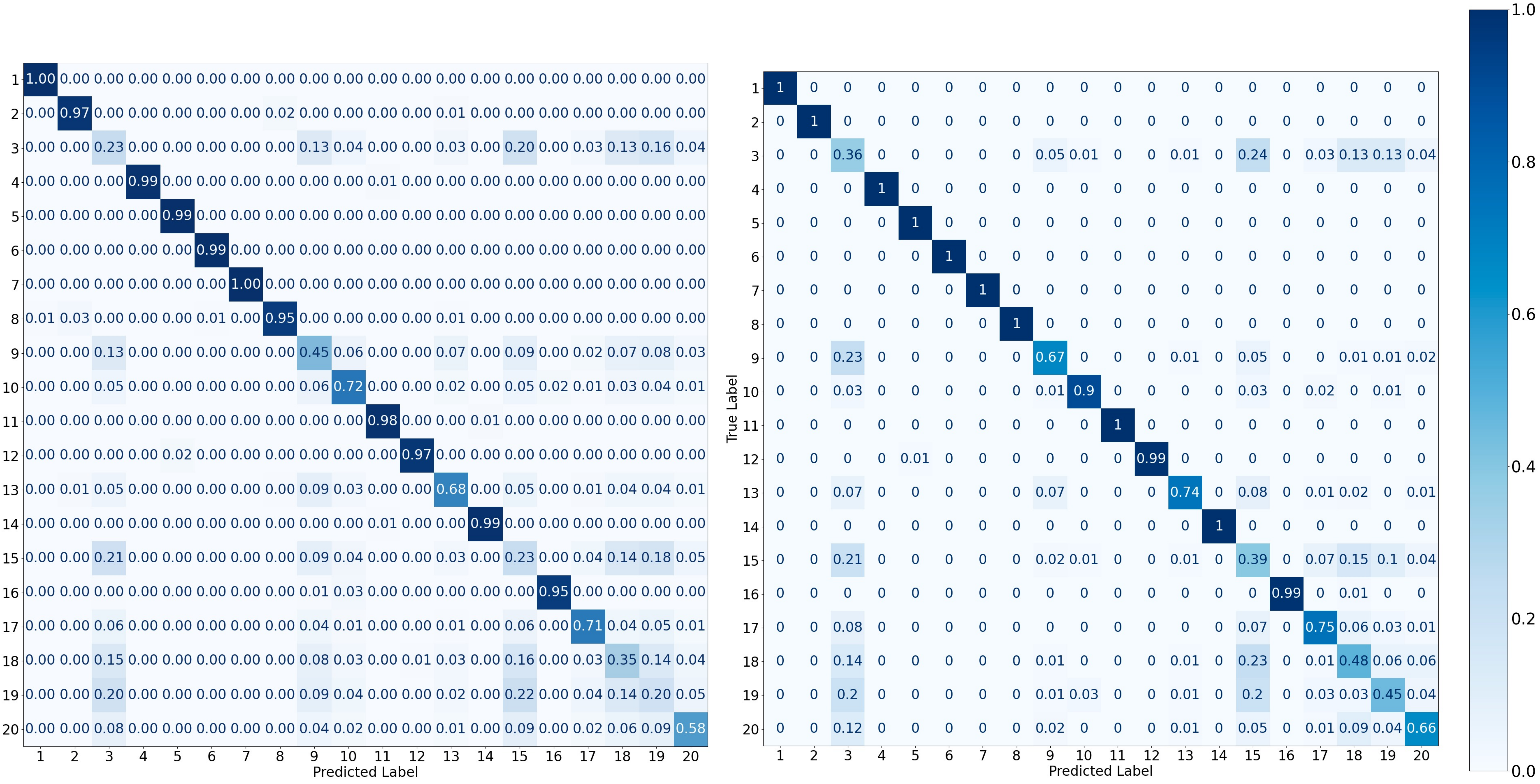}
    \caption{Our modified SVM method outperforms supervised LSTM AE model fault classification accuracy. (Left): The confusion matrix of supervised LSTM AE model with 74\% fault detection accuracy; (Right): Confusion matrix of SVM method trained on mapped covariance matrices shows an improved classification accuracy of 82\%. The patience time is set to 700 samples for both cases.}
    \label{fig: compare_lstm_SAFIRE}
\end{figure}

\begin{figure}[!ht]
    \centering
    \includegraphics[width=0.6\textwidth]{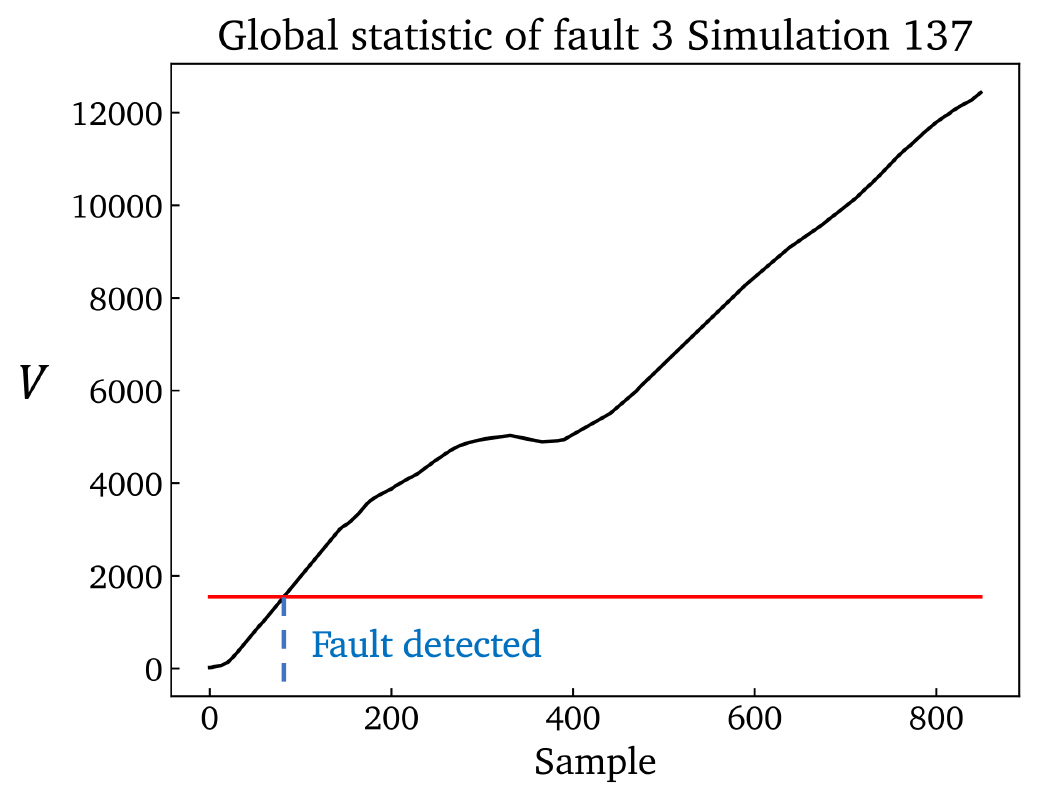}
    \caption{Global statistic ($V(t)$) plot for Fault 3 in one of the simulations with $t_p = 700$. The key features extracted for this plot are mean $= 6200.16$, standard deviation $= 3351.48$, median $= 5331.36$, variance $= 11232431.32$, range $= 12416.88$, max $= 12436.18$, AUC $=5263904.62$, and the number of peaks $= 3$. }
    \label{fig: example_global_stat}
\end{figure}

\begin{figure}[!ht]
    \centering
    \includegraphics[width=0.55\textwidth]{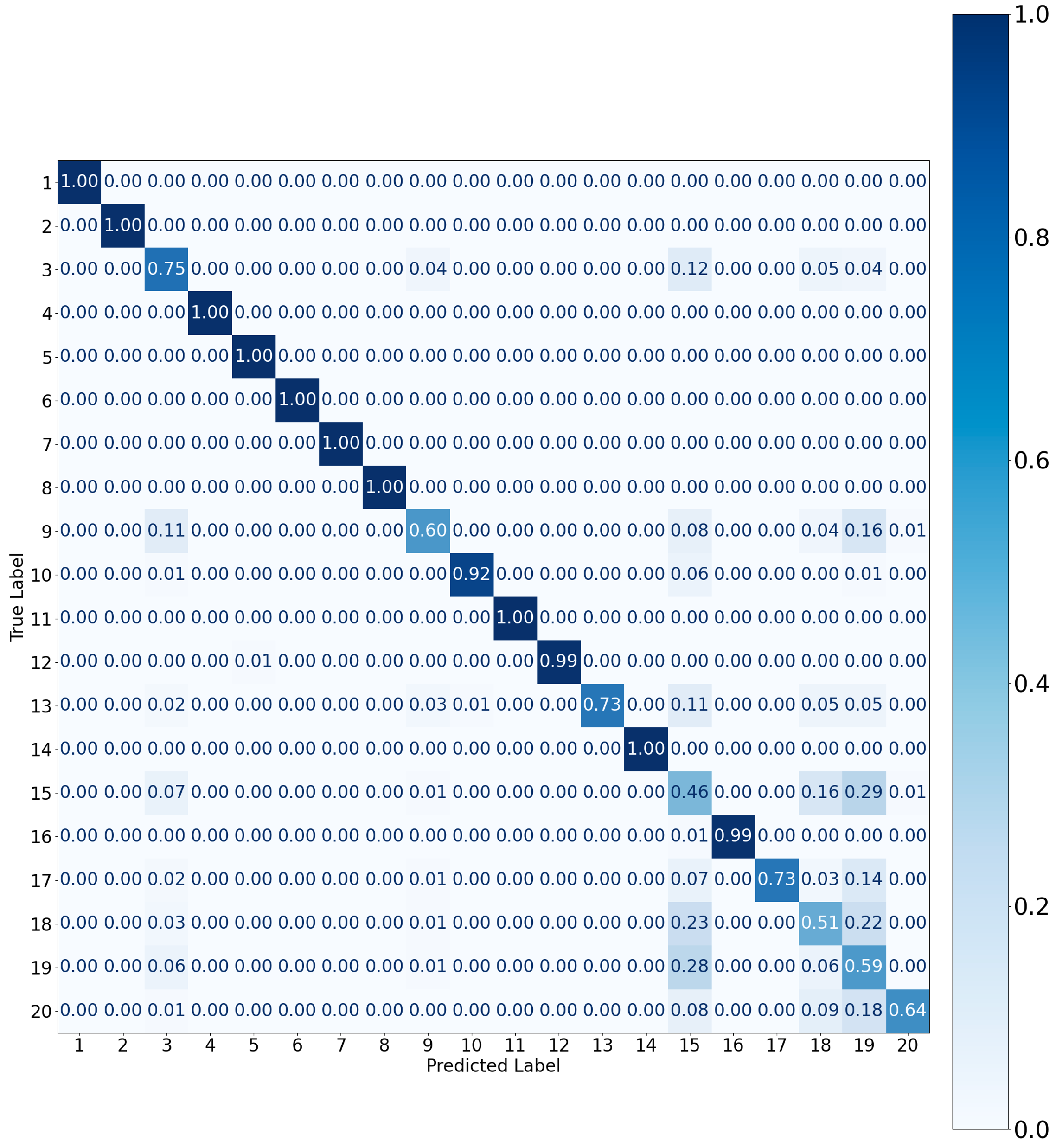}
    \caption{Confusion matrix of modified SVM method shows an improved fault classification accuracy of 84.5\% when introducing global statistics features from eCDF-SPC method into the feature set.}
    \label{fig: conf_mat_improved}
\end{figure}

Lastly, to further improve the classification accuracy, we extract the statistical features from global statistic $V(t)$ plot for each simulation. Figure \ref{fig: example_global_stat} shows an example of the global statistic $V(t)$ plot for Fault 3 in one of simulations. As a proof-of-concept study, we extract key metrics such as mean, standard deviation, median, variance, range, maximum value, number of peaks, and area under the curve (AUC) using \texttt{Numpy} \cite{numpy} and \texttt{SciPy} libraries in Python \cite{scipy}. By incorporating these statistical features obtained from the fault detection module into fault classification, we observe even further improvement in fault classification accuracy from 82\% to 84.5\% (see confusion matrix in Figure \ref{fig: conf_mat_improved}), thus demonstrating the synergistic enhancement in process monitoring performance by integrating fault detection and classification in a holistic framework such as FARM.

\section{Conclusion}
In this study, we enhance our FARM process monitoring framework presented at the FOCAPD 2024 Conference \cite{alireza:2024:focapd} by introducing the latest advancement in anomaly detection of nonparametric, heterogeneous big data streams using eCDF-based SPC method. Furthermore, our comprehensive case study on the TEP benchmark problem indicates that by this new SPC framework achieves outstanding performance and a good balance among speed, accuracy, and robustness. In terms of fault classification, by introducing basic concepts in Riemannian geometric analysis to well established classification methods, such as the SVM, we observe significant improvement in fault detection accuracy. Furthermore, we show that by incorporating selected statistical features from fault detection module, fault classification performance can be further improved. These promising results open up many exciting opportunities for future research in chemical process monitoring.

\section*{Acknowledgments}
A.M. and Z.J. acknowledge financial support from U.S. National Science Foundation (NSF) award number 2331080. F.M. and Z.J. acknowledge financial support from Oklahoma Center for Advancement of Science and Technology (OCAST) Oklahoma Applied Research Support (OARS) program grant number AR24-069. Financial support from the startup fund of College of Engineering, Architecture, and Technology at Oklahoma State University is also greatly acknowledged.

\bibliographystyle{plainnat}  
\bibliography{references}

\end{document}